\documentclass[fleqn,10pt]{wlscirep}
\usepackage[utf8]{inputenc}
\usepackage[T1]{fontenc}
\usepackage[squaren]{SIunits}
\usepackage{bm}
\usepackage{lineno}
\title{Superradiant Thomson Scattering via Oscillating Quasiparticles}

\author[1,*]{Qianyi Ma}
\author[1]{Yuhui Xia}
\author[1]{Zhenan Wang}
\author[1]{Letian Liu}
\author[1]{Zhiyan Yang}
\author[1,3,*]{Xinlu Xu}
\author[1,3,4,*]{Xueqing Yan}
\affil[1]{State Key Laboratory of Nuclear Physics and Technology, and Key Laboratory of HEDP of the Ministry of Education, CAPT, School of Physics, Peking University, Beijing 100871, China}
\affil[2]{Beijing Laser Acceleration Innovation Center, Huairou, Beijing, 101400, China}
\affil[3]{Institute of Guangdong Laser Plasma Technology, Baiyun, Guangzhou, 510540, China}

\affil[*]{xuxinlu@pku.edu.cn;x.yan@pku.edu.cn}

\begin{abstract}
The recently proposed concept of generalized superradiance (GS) allows for the generation of coherent radiation without the need for complex compression or prebunching of relativistic electrons. However, the quasiparticles in current GS schemes are formed through local electron accumulation and their sizes exceed 100 nm, restricting the achievable radiation wavelength. In this study, we demonstrate a realization of narrow quasiparticles with $\lesssim$10 nm widths through sheet-crossing. When an energy chirped electron beam collides with an intense laser pulse, lower energy electrons at the front slip back, forming an accelerating quasiparticle. These quasiparticles undergo transverse oscillations within the laser field, thereby extending GS emission from the Cherenkov regime into the synchrotron regime. Three-dimensional particle-in-cell simulations indicate the production of gigawatt-class chirped radiation in the 10-100 nm range. The proposed scheme is compatible with existing Thomson scattering facilities, paving the way for the generation of coherent ultrafast radiation.
\end{abstract}
\begin{document}

\flushbottom
\maketitle
%
%
\thispagestyle{empty}


Relativistic electron beams produced by conventional radio-frequency (RF) accelerators \cite{chao2020lectures} or plasma-based accelerators (PBAs) \cite{1979lwfa} can emit radiation across a broad spectrum, providing a strong complement to atomic lasers by extending coverage into spectral regions that lasers cannot readily access \cite{2017synchrotron-xrays, 2013PBA-xrays-review, Emma2010NP_LCLS, pellegrini2016physics, 2021wang-fel, Fisher2022NP, Kang2026NP}. If the beam length is shorter than the radiation wavelength \cite{2007flying-mirror,2010wu-flying-mirror} or it consists of beamlets with a spacing equal to the radiation wavelength \cite{li2008nonrelativistic, neumann2009terahertz, PhysRevLett.101.054801, PhysRevLett.105.234801, liang2023widely, xu2016nanoscale, PhysRevLett.125.014801, xu2022generation}, the emission transitions from incoherent spontaneous radiation to coherent radiation, with orders-of-magnitude intensity enhancement and significant spectral narrowing \cite{2019review-superradiance}. This coherent or superradiant regime lies at the heart of modern accelerator-based light sources such as free-electron lasers (FELs) \cite{Emma2010NP_LCLS, pellegrini2016physics, 2016LCLS-review, 2017xfel-sources-review, ROSSBACH2019, 2021wang-fel, Fisher2022NP, Kang2026NP, Liang2026LSA}.

However, it is challenging to compress or prebunch electron beams at extreme ultraviolet or x-ray wavelengths to achieve superradiant emission due to the repulsive force between electrons. Alternatively, Ref. \cite{2021generalized-superradiance} generalizes the concept of superradiance by organizing the motion of the electrons to form an apparent quasiparticle. Unlike conventional microbunched or ultrashort beams composed of a fixed set of constituent electrons, the quasiparticle manifests as an apparent density or current structure that radiates as a single coherent entity, sustained by a continuous flux of individual electrons entering and exiting the structure. It was realized in simulations by driving a nonlinear plasma wake in a density upramp, whose high-density tail moves superluminally and produces Cherenkov radiation superradiantly \cite{2023cherenkov-quasiparticle, 2024cherenkov-quasiparticle}. Another simulation realization of superluminal quasiparticles was achieved using spirally modulated rotating electrons \cite{2021generalized-superradiance} generated by shooting a circularly polarized (CP) intense laser into a microtube \cite{2025spp-quasiparticle}. The generalized superradiance (GS) holds great potential for the realization of compact and coherent light sources.

While the electron beam itself can be much longer, the longitudinal thickness of the quasiparticle in GS must be comparable to, or shorter than, the radiation wavelength to achieve superradiance, consistent with the requirements in conventional superradiance. In the scenarios realized so far, the effective quasiparticle thickness is hundreds of nanometers; consequently, the GS spectrum is restricted to wavelengths greater than 100 nm \cite{2023cherenkov-quasiparticle,2024cherenkov-quasiparticle,2025spp-quasiparticle}. Furthermore, these quasiparticles move superluminally in the longitudinal direction to emit Cherenkov radiation. In this work, we present a new realization of quasiparticles in Thomson scattering configuration where relativistic electron beams collide with intense laser pulses \cite{PhysRevLett.10.75, ARUTYUNIAN1963, Sandorfi1983, 1992lss-source, Ting1995JAP, PhysRevLett.76.3116}. By exploiting the energy chirp of the electron beam, narrow quasiparticles with a width of $\lesssim$10 nm are formed via sheet crossing. We also identify another class of quasiparticles, formed via local electron accumulation. Apart from their longitudinal motion, these quasiparticles undergo transverse oscillations inside the laser pulse, which extend the GS emission from Cherenkov radiation to synchrotron radiation. The proposed scheme can be realized in existing Thomson scattering facilities \cite{Schoenlein1996Science, Gibson2004PLEIADES, WELLER2009HIgammaS, TTX2013RSI, FukudaIPAC2016-TUPOW046, PhysRevAccelBeams.22.053403, Wang2022SLEGS, SAMSAM2024STAR,  PhysRevAccelBeams.28.023401, Gunther2023, Lyncean2017, LumitronTechnologies, PhysRevLett.96.014802, ta2012all, powers2014quasi, PhysRevLett.114.195003, yan2017high, PhysRevX.8.011020, PhysRevX.8.031004, Ma2020MRE, mirzaie2024all, matheron2024comptonphotonsgevscale, hpl2025sjtu-ts} to significantly enhance their capability by generating coherent ultrafast x-rays. This represents a major step in advancing GS from a concept to a practically valuable realization.

\section*{Results}

\begin{figure}[htbp]
\includegraphics[width=1.0\linewidth]{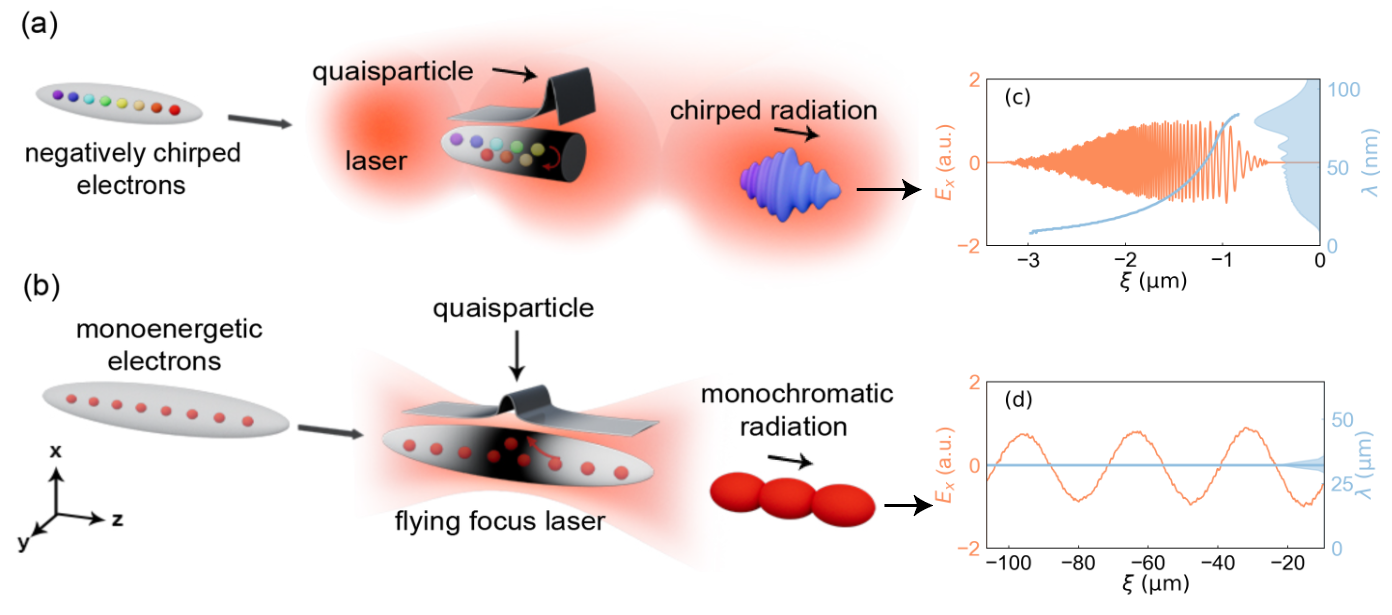}
\caption{\label{fig:concept} Schematic configurations for the formation of the first (a) and second (b) classes of quasiparticles in a head-on collision between an electron beam and a laser pulse, and the corresponding radiation fields (orange lines), spectrum (blue shaded regions) and central frequencies (blue lines) for the first (c) and second (d) classes. Colored dots in (a) and (b) represent electrons with different energies, with red/blue indicating low/high energy. }
\end{figure}

As illustrated in Fig. \ref{fig:concept}(a), a negatively chirped electron beam collides with a counter-propagating laser pulse. As an electron enters the laser, it oscillates transversely, causing a reduction in the longitudinal velocity ($v_z$). The conservation of the canonical momentum of an electron gives $v_z \approx (1-\frac{1+a_\mathrm{L}^2}{2\gamma_0^2})c$, where $\gamma_0$ is the initial relativistic factor of the electron, $a_\mathrm{L}=\frac{eE_\mathrm{L}}{m_\mathrm{e}\omega_\mathrm{L} c}$ is the normalized vector potential of the laser, $e$ and $m_e$ are the electron charge and mass, $c$ is the speed of light in vacuum, $E_\mathrm{L}$ and $\omega_\mathrm{L}=\frac{2\pi c}{\lambda_\mathrm{L}}$ are the electric field and frequency of the laser pulse. A CP laser pulse is assumed throughout this work for simplicity; a linearly polarized (LP) laser yields similar results, except that it introduces a factor of $\frac{1}{2}$ to the $a_\mathrm{L}^2$ term and may generate higher-order harmonics (see Supplemental Material). The laser enhances the velocity difference between electrons with different energies by a factor of $a_\mathrm{L}^2$. Electrons with lower energies slip backward faster than the higher-energy electrons behind, leading to sheet crossing and formation of a narrow density spike. This density spike oscillates transversely and radiates as a quasiparticle, which sweeps from the beam head (with the lowest energy) to the tail (with the highest energy). Electron sheet crossing ensures a narrow and high-density quasiparticle. 

In Thomson scattering, quasiparticles can also be formed via gathering of monoenergetic electrons. Since the laser intensity is high in the vicinity of its focus, electrons near the focus have a markedly reduced longitudinal velocity, causing them to slip backward toward the electrons behind them and form a density spike [Fig. \ref{fig:concept}(b)]. The velocity of this quasiparticle is equal to the velocity of the laser intensity peak, which can be arbitrary for structured pulses \cite{2017flyfoc-concept}. This quasiparticle is formed by electron gathering as in \cite{2021generalized-superradiance, 2023cherenkov-quasiparticle, 2024cherenkov-quasiparticle, 2025spp-quasiparticle}, and is therefore much wider than the one formed via sheet crossing.

\hspace*{\fill}

\noindent\textbf{Dynamical behavior of the two classes of quasiparticles. }
Consider two electrons with initial coordinates $\xi_0  -\frac{\Delta \xi_0}{2}$ and $\xi_0+\frac{\Delta \xi_0}{2}$, where $\xi\equiv z -c t$ and the subscript `0' represents the initial quantities, and we assume $\Delta \xi_0 $ is much shorter than the characteristic length of the laser envelope. The initial coordinate affects the electron velocity at time $t$ in two ways: by altering the initial energy and by determining the local $a_\mathrm{L}$ experienced by the electron. The velocity difference between these two electrons at time $t$ thus contains two terms as $\Delta v_z =  \frac{\partial v_z}{\partial \gamma_0}\frac{\mathrm{d} \gamma_0}{\mathrm{d} \xi_0}\Delta \xi_0  + \frac{\partial v_z}{\partial a_\mathrm{L}} \frac{\partial a_\mathrm{L}}{\partial \xi_0}\Delta \xi_0$. Correspondingly, the distance between electrons evolves as $\Delta \xi = \Delta \xi_0 +  \Delta \xi_0 \int_0^t \mathrm{d}t^{\prime} \left[ \frac{\partial v_z}{\partial \gamma_0}\frac{\mathrm{d} \gamma_0}{\mathrm{d} \xi_0} + \frac{\partial v_z}{\partial a_\mathrm{L}} \frac{\partial a_\mathrm{L}}{\partial \xi_0}\right] $. 

For electron beams with an energy chirp and lasers with slowly varying envelope, we keep only the first term and have $\Delta \xi \approx \Delta \xi_0 + \Delta \xi_0\int_0^t \mathrm{d}t^{\prime} \frac{\partial v_z}{\partial \gamma_0}\frac{\mathrm{d}\gamma_0}{\mathrm{d}\xi_0}$. The coincidence condition for these two electrons ($\Delta \xi=0$) gives $\int_0^t \mathrm{d}t^{\prime} \frac{ \partial v_z} {\partial \gamma_0}\frac{\mathrm{d}\gamma_0}{\mathrm{d}\xi_0} = \frac{\mathrm{d} \gamma_0 / \mathrm{d} \xi_0}{\gamma_0^3}\int_0^t \mathrm{d}t^{\prime} \left(1+a_\mathrm{L}^2 \right) \approx  - 1 $. This indicates that the constituent electrons of the quasiparticle change with time and they can be denoted as $\xi_0(t)$ or $\gamma_0(t)$. For a linear chirp of $k\equiv \frac{\mathrm{d}\gamma_0}{\mathrm{d}\xi_0}<0$ and a flattop laser envelope, we have $\gamma_0(t) = \left[ - k (1+a_\mathrm{L}^2)ct\right]^{1/3}$, i.e., lower-energy electrons constitute the quasiparticle first, followed by higher-energy electrons. Here the beam is initialized with $\gamma_0=k\xi_0$, and the laser propagating along $-z$ is launched from $z=0$ at $t=0$. For nonlinear chirps and varying laser envelopes, $\gamma_0(t)$ can be given numerically and controlled by the chirps and envelopes. 

The first-class quasiparticle appears at $\xi^\mathrm{qI}=\xi_0(t)+\int_0^t \mathrm{d}t^{\prime} \left[ v_z(\gamma_0(t)) - c \right]$ and its forward velocity is $v_{z}^{\mathrm{qI}}= v_z(\gamma_0(t)) + \frac{\mathrm{d}\xi_0}{\mathrm{d}t} \left[ 1 + \int_0^t \mathrm{d}t^{\prime}  \frac{\partial v_z(\gamma_0(t))}{\partial \gamma_0}\frac{\mathrm{d}\gamma_0}{\mathrm{d} \xi_0} \right]$. Applying the coincidence condition, we obtain $v_{z}^{\mathrm{qI}} \approx v_z (\gamma_0(t))$, i.e., the quasiparticle's longitudinal velocity is equal to that of its constituent electrons and therefore satisfies $0<v_{z}^{\mathrm{qI}}<c$. For a linear chirp and a flattop envelope, the quasiparticle has a forward relativistic factor as $\gamma_{z}^{\mathrm{qI}} \equiv \frac{1}{\sqrt{1-(v_{z}^{\mathrm{qI}}/c)^2}} = \frac{\left[ - k (1+a_\mathrm{L}^2)ct\right]^{1/3}}{\sqrt{1+a_\mathrm{L}^2}}$, indicating it accelerates. 

For two electrons with the same energy, they share the same trajectory except a temporal delay and thus cannot coincide.  We keep only the second term in $\Delta v_z$ and have $\Delta \xi \approx \Delta \xi_0 \left( 1 + \int_0^t \mathrm{d}t^{\prime} \frac{\partial v_z}{\partial a_\mathrm{L}} \frac{\partial a_\mathrm{L}}{\partial \xi_0} \right) \approx \Delta \xi_0 \left( 1- \frac{1}{2\gamma_0^2}\int_0^t \mathrm{d}t^{\prime}  \frac{\partial a_\mathrm{L}^2}{\partial \xi_0} \right)$. This indicates that $\Delta \xi$ decreases on the rising edge of the laser pulse and increases on the falling edge; it reaches a minimum at the intensity peak, leading to the formation of a density peak. The motion of this second-class quasiparticle is governed by the spatial-temporal distribution of the laser pulse, which can be controlled for a flying-focus laser whose intensity peak is always at its focus \cite{2017flyfoc-concept}. The forward velocity of the second-class quasiparticle is $v_z^\mathrm{qII}=v_\mathrm{f}$, where $v_\mathrm{f}$ is the focus velocity of the laser. Its velocity is decoupled from that of the constituent electrons and can be arbitrary (even superluminal as in previous work \cite{2021generalized-superradiance, 2023cherenkov-quasiparticle, 2024cherenkov-quasiparticle, 2025spp-quasiparticle}).

\begin{figure}[!htbp]
\centering
\includegraphics[width=0.75\linewidth]{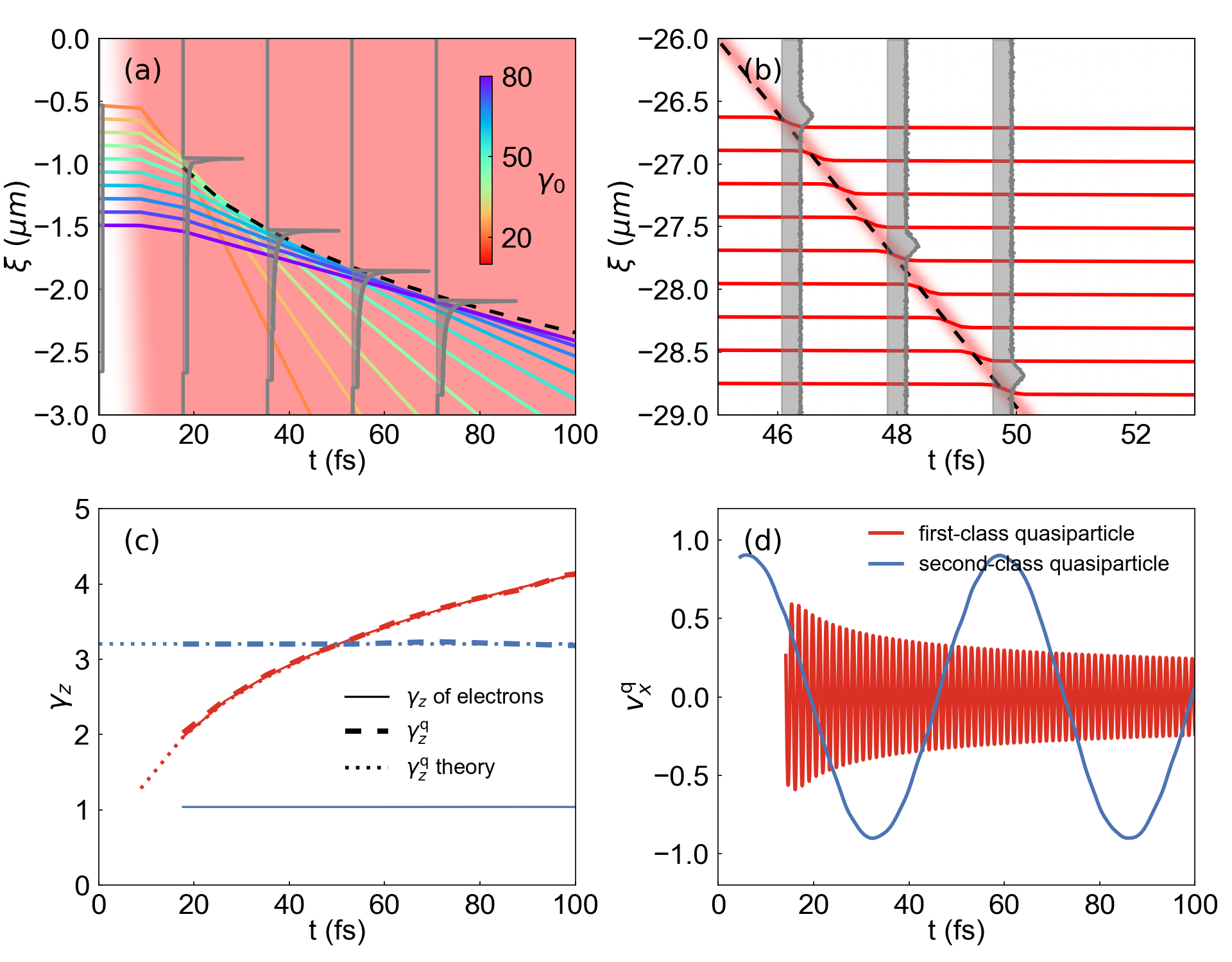}
\caption{\label{fig:dynamics} The dynamic behaviors of the quasiparticles. Trajectories of electrons in $t-\xi$ for the first class (a) and the second class (b) of quasiparticles. The black dashed lines indicate the predicted quasiparticle trajectory, and the gray and red shaded areas denote the electron density distribution and the laser envelope. (c) The evolution of $\gamma_z$ of the quasiparticles (dashed lines) and their constituent electrons (solid lines). The dotted lines represent the theoretical predictions. (d) The evolution of the transverse velocity of the quasiparticles. In (c) and (d), the red and blue lines denote the first and second classes of quasiparticles, respectively. }
\end{figure}

We perform one-dimensional (1D) simulations using the fully relativistic PIC code OSIRIS \cite{osiris} to verify the quasiparticle formation. The detailed setup can be found in the Supplemental Material. In Fig. \ref{fig:dynamics}(a), a $2.1~\micro\meter$-long electron beam with a negative energy chirp $k=-37.6~\micro\meter^{-1}$ collides with an 800-nm CP laser pulse with a sharp rising edge followed by a flattop envelope and $a_0=15$, where $\gamma_0=60$ at the beam center and the rms slice energy spread is $\sigma_{\gamma_0}\equiv\sqrt{\langle\gamma_0^2\rangle_\xi-\langle\gamma_0\rangle_\xi^2}=0.1$, with $\langle  \rangle_\xi$ denoting the ensemble average over electrons at a slice $\xi$. The needed energy chirp can be produced in PBAs \cite{emma2021terawatt, 2024xu-self-selection, ma2025generating} since the acceleration field in the nonlinear plasma wake has a natural chirp as $\frac{\mathrm{d}E_z}{\mathrm{d}z}\approx -\frac{0.1~\giga\volt\per\meter\meter}{\micro\meter}$ for typical $10^{19}~\centi\meter^{-3}$ plasma densities \cite{PhysRevLett.96.165002}. As shown in Fig. \ref{fig:dynamics}(a), the electrons encounter the laser leading edge after a drift of $\sim10~\femto\second$ and then slip backward notably in $\xi$. Sheet crossing occurs when higher-energy electrons in the back catch up with lower-energy electrons in the front [intersections of adjacent trajectories in Fig. \ref{fig:dynamics}(a)], thereby forming a narrow density spike as shown by the gray lines. The density spike (quasiparticle) slips back gradually, and its longitudinal motion agrees well with the theory (black dashed line). 

For the second class of quasiparticles, as shown in Fig. \ref{fig:dynamics}(b), a $106~\micro\meter$-long monoenergetic electron beam with $\gamma_0=10$ collides with an 800-nm CP flying-focus laser pulse with $a_0=15$, where the focal point travels at $v_\mathrm{f}=-0.95c$. The flying focus laser has a Gaussian temporal envelope with a duration of $0.42$ fs (see Supplemental Material). This electron beam can be readily delivered by conventional RF accelerators. The electrons drift in vacuum except near the laser focus, where notable deceleration occurs, as shown by the red lines in Fig. \ref{fig:dynamics}(b). The aggregation of the electrons forms a density peak as shown by the gray lines. This quasiparticle propagates backward at the laser focus velocity (black dashed line). The width of this quasiparticle is determined by the laser envelope width, $\sim 100~\nano\meter$, and is significantly wider than the first-class quasiparticle ($\sim 10~\nano\meter$). Meanwhile, the quasiparticle's density is only twice the initial beam density, whereas that of the first-class quasiparticle is an order of magnitude higher. 

Figure \ref{fig:dynamics}(c) shows the evolution of $\gamma_z^\mathrm{q}$. The first-class quasiparticle is accelerated as shown by the red dashed line which agrees well with the formula (red dotted line). For the second-class quasiparticle, $\gamma_z^\mathrm{q}$ remains constant at 3.2. We also plot the $\gamma_z$ of electrons constituting the quasiparticles (solid lines), which matches with $\gamma_z^{\mathrm{q}}$ for the first class and differs from $\gamma_z^{\mathrm{q}}$ for the second class. The supplement shows more examples of the quasiparticles, including the superluminal second-class quasiparticles.

In contrast to previous work where the quasiparticle undergoes uniform motion along a straight line \cite{2021generalized-superradiance,2023cherenkov-quasiparticle,2024cherenkov-quasiparticle,2025spp-quasiparticle}, the quasiparticles proposed here exhibit transverse oscillations with a frequency $(1+\frac{v_z^\mathrm{q}}{c})\omega_\mathrm{L}$ while propagating forward with $v_z^\mathrm{q}$. This is because the thickness of the quasiparticles is much narrower than the laser wavelength, its constituent electrons have similar oscillation phases, and their transverse velocities add coherently. Figure \ref{fig:dynamics}(d) shows the oscillation of the mean transverse velocity of electrons constituting the quasiparticle, whose frequency is $\omega^\mathrm{qI}\approx 2 \omega_\mathrm{L}$ (red line) and $\omega^\mathrm{qII}\approx 0.05 \omega_\mathrm{L}$ (blue line). For the first class, the amplitude of $v_x^\mathrm{qI}$ decreases continuously due to the increase of $\gamma_z^\mathrm{qI}$, while for the second class, the amplitude remains constant due to the constant $\gamma_z^\mathrm{qII}$. As discussed below, the oscillating quasiparticles enable generalized superradiant radiation beyond the Cherenkov radiation emitted by non-oscillating quasiparticles \cite{2023cherenkov-quasiparticle,2024cherenkov-quasiparticle,2025spp-quasiparticle}. 

\hspace*{\fill}

\begin{figure}[!htbp]
\centering
\includegraphics[width=\linewidth]{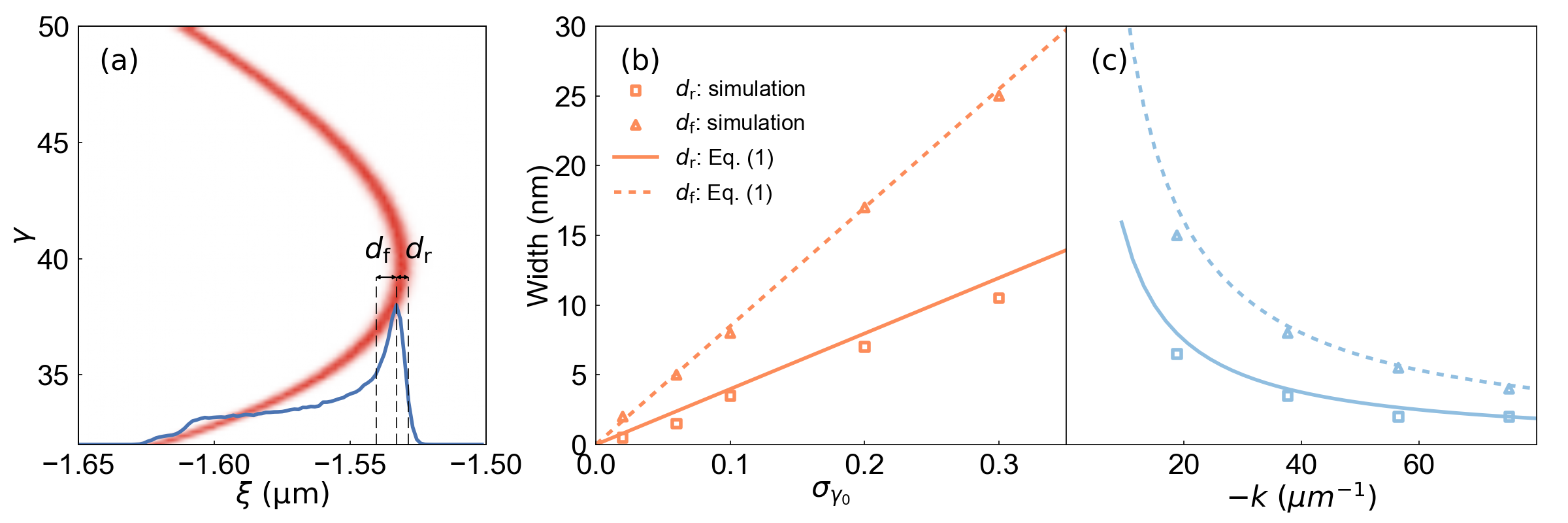}
\caption{\label{fig:width} (a) The longitudinal phase space distribution of the electron beam at $t=35$ fs for the first-class quasiparticle. The blue line is the density profile. Dependence of $d_\mathrm{r}$ and $d_\mathrm{f}$ on $\sigma_{\gamma_0}$ (b) and $k$ (c). }
\end{figure}

\noindent\textbf{Longitudinal thickness of the quasiparticles.} The thickness of the quasiparticle imposes a limit on the minimum wavelength of the superradiant radiation, and thus is a key parameter. For the second-class quasiparticle, its thickness is approximately equal to the temporal width of the laser envelope (see simulation confirmation in the supplement). We show the longitudinal phase space distribution and the density profile for the first-class quasiparticle in Fig. \ref{fig:width}(a), where sheet crossing occurs for electrons with $\gamma_0=40$. The quasiparticle has a sharp rising edge and a slow falling edge. If the initial slice energy spread is $\sigma_{\gamma_0}=0$, sheet crossing causes a singularity with a falling edge as $\propto \frac{1}{\sqrt{\eta}}$, where $\eta$ is the distance to the singularity. When $\sigma_{\gamma_0}\neq 0$, the beam can be decomposed into cold chirped beamlets with slightly different energies and they form their own density spike at slightly different locations. The overall density profile can thus be obtained by integrating over the energy distribution (see detailed derivations in the supplement). For a Gaussian slice energy distribution, the half-width at half-maximum on the rising and falling edges is given by
\begin{align}
    d_\mathrm{r}\approx -1.5\frac{\sigma_{\gamma_0}}{k}, d_\mathrm{f}\approx -3.2\frac{\sigma_{\gamma_0}}{k}. \label{eq:d_width}
\end{align}

Figures \ref{fig:width}(b) and (c) show the dependence of $d_\mathrm{r}$ and $d_\mathrm{f}$ on $\sigma_{\gamma_0}$ and $k$, respectively. The simulation results for $d_\mathrm{r}$ (squares) and $d_\mathrm{f}$ (triangles) agree well with Eq. \eqref{eq:d_width} (lines). A small $\sigma_{\gamma_0}$ and a large $|k|$ lead to a thin quasiparticle. Density downramp injection in PBAs can produce high-quality electron beams with $\sigma_{\gamma_0}\sim 0.1$ and tunable $k$ \cite{xu2017downramp}, which can form quasiparticles with 1-10 nm width, orders of magnitude narrower than those in other works \cite{2023cherenkov-quasiparticle,2024cherenkov-quasiparticle,2025spp-quasiparticle}. Ionization injection enabled by transversely colliding laser pulses can generate bright electron beams with $\sigma_{\gamma_0} \approx 0.02$ \cite{Li2013PhysRevLett.111.015003}, which can reduce the width to $\sim$nm. Operating PBA at solid plasma densities can significantly improve the gradient of the acceleration field to increase $|k|$ and compress quasiparticles to angstrom-scale width \cite{ma2025generating}. 

\hspace*{\fill}

\noindent\textbf{Superradiant quasiparticle radiation. }Using the concept of quasiparticles, the radiating current density can be expressed as $\bm{j}(\bm{r},t)=ef[\bm{\zeta^\mathrm{q}}(t)]\left[ v_z^\mathrm{q}(t)\bm{z} + \bm{v}_\perp^\mathrm{q}(t)\right]$, where $\bm{\zeta}^\mathrm{q}\equiv \bm{r}-\bm{r}^\mathrm{q}(t)$, $\bm{r}^\mathrm{q}(t)$ is the trajectory of the quasiparticle, and $f(\bm{\zeta^\mathrm{q}})$ represents the 3D density distribution of the quasiparticle. Substituting it into the angular spectrum distribution of the far-field radiation intensity \cite{book:jackson}, we have 
\begin{equation}
    \frac{\mathrm{d}^2 I}{\mathrm{d}\omega \mathrm{d}\Omega}\approx \frac{e^2\omega^2}{16\pi^3 \epsilon_0 c^3}\biggl|\int_{-\infty}^\infty \mathrm{d}t g(\omega,\Omega)\biggl[v_z^\mathrm{q} \bm{n}\times(\bm{n}\times \bm{z}) + \bm{n}\times(\bm{n}\times \bm{v}_\perp^\mathrm{q})  \biggr] \mathrm{e}^{i\omega(t-\frac{\bm{n}\cdot\bm{r}^\mathrm{q}}{c})} \biggr|^2, \label{eq:I}
\end{equation}
where $\bm{n}$ is a unit vector in the radiation propagation direction, $g(\omega,\Omega)=\int_{-\infty}^\infty \mathrm{d}\bm{\zeta}^{\mathrm{q}} f(\bm{\zeta}^\mathrm{q})\mathrm{e}^{-i\omega \frac{\bm{n}\cdot\bm{\zeta}^\mathrm{q}}{c}}$ is the 3D shape factor of the quasiparticle. The two terms in Eq. (2) reflect a fundamental transition in the underlying radiation mechanisms. In previous GS, the quasiparticle moves strictly longitudinally, emitting Cherenkov-type radiation when its velocity exceeds the velocity of light ($|v_z^\mathrm{q}|>c$) \cite{2023cherenkov-quasiparticle,2024cherenkov-quasiparticle,2025spp-quasiparticle}. In our configuration, however, the quasiparticle undergoes a coherent transverse oscillation. This transverse acceleration is physically analogous to a relativistic charged particle wiggling in a synchrotron or undulator, thereby enabling a transition to an unexplored regime of quasiparticle synchrotron-like radiation.

We focus on the term with $\bm{v}_\perp^\mathrm{q}$ in Eq. \eqref{eq:I}. For simplicity, we approximate the quasiparticle profile as a Gaussian function with a rms width of $d=0.4(d_\mathrm{r}+d_\mathrm{f})$ and use $\bm{v}_\perp^\mathrm{q}=\frac{v_{\perp 0}^\mathrm{q}}{\sqrt{2}}(\bm{x}+i\bm{y})\mathrm{e}^{i\omega^\mathrm{q}t}$ to obtain
\begin{equation}
    \frac{\mathrm{d}^2 I}{\mathrm{d}\omega \mathrm{d}\Omega}\Big|_{\theta=0} \approx \frac{e^2\omega^2(v_{\perp0}^\mathrm{q}\gamma_z^\mathrm{q})^2 }{4\pi\epsilon_0 c^3}N^2 \mathrm{e}^{-\frac{\omega^2d^2}{c^2}} \delta^2\left(\omega-\omega_\mathrm{r}\right), \label{eq:I}
\end{equation}
where $\omega_\mathrm{r}=\omega_\mathrm{L}\frac{1+v_z^\mathrm{q}/c}{1-v_z^\mathrm{q}/c}$ is the radiation frequency, the same as the Thomson scattering \cite{book:jackson}. Since a CP laser is assumed here, no harmonic component is present in Eq.~\eqref{eq:I}. If an LP laser were used, the first-class quasiparticle would undergo a longitudinal velocity oscillation and emit higher-order harmonics, while the second-class quasiparticle would maintain a constant longitudinal velocity strictly locked to the focal velocity of the laser, thereby generating no harmonics. The coherent enhancement factor over the incoherent $N$-particle radiation is $Ne^{-\frac{\omega^2d^2}{c^2}}$, which is much larger than 1 when $\omega \lesssim \frac{c}{d}$. 

As shown Fig. 1(c), a chirped radiation pulse with wavelengths ranging from 8 nm to 90 nm is formed due to the accelerated first-class quasiparticle. The radiation intensity decreases toward the end of the emission, as the shortening wavelength reduces the shape factor. For the second class [Fig. \ref{fig:concept}(d)], a monochromatic radiation pulse with a wavelength of $\lambda_\mathrm{r} \approx 31.2~\micro\meter$ is produced, which agrees well with the prediction $\lambda_\mathrm{L}\frac{1+v_z^\mathrm{qII}/c}{1-v_z^\mathrm{qII}/c}=31.2~\micro\meter$.

We further perform 3D simulations with an advanced Maxwell solver \cite{xu-solver} to study high-dimensional effects on the first-class quasiparticle radiation with high-fidelity \cite{xu2013numerical}. The electron beam has a Gaussian transverse profile (rms spot size $\sigma_{r0}=1.0~\micro\meter$), a flattop longitudinal profile (length $1.6~\micro\meter$), a normalized emittance of $\epsilon_\mathrm{n}=106$ nm, an initial peak density of $10^{20}~\mathrm{cm}^{-3}$ and a chirp of $k=-56.4~\micro\meter^{-1}$. We employ a flying focus laser with a spot size $w_0=4.2~\micro\meter$, a peak power of $121~\mathrm{TW}$ ($a_\mathrm{L}=10$), and a focal velocity of $0.95c$. Unlike its role for second-class quasiparticles, the flying-focus configuration here is utilized solely to maintain the short electron beam near the laser intensity peak, thereby reducing the required pulse energy. The quasiparticle itself is generated via sheet crossing, maintaining a width much narrower than the laser intensity envelope. Other parameters remain the same as in the 1D case.

\begin{figure}[!htbp]
\centering
\includegraphics[width=0.75\linewidth]{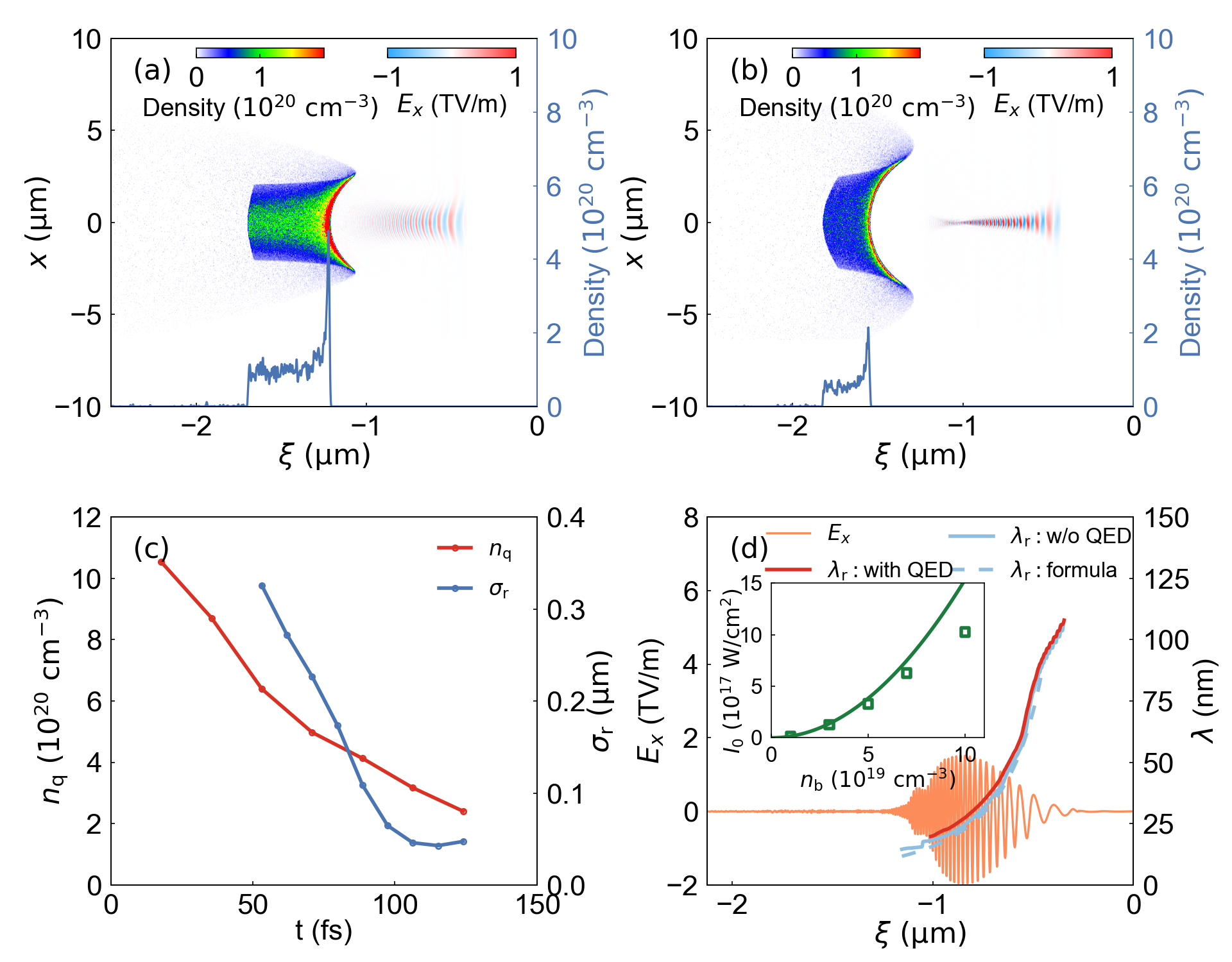}
\caption{\label{fig:3} 3D simulation of the first-class quasiparticle. Electron density and $E_x$ field in the $y=0$ plane at (a) $t=53$ fs and (b) $t=106$ fs. The blue curves show the on-axis beam density profile. (c) The evolution of the quasiparticle density $n_\mathrm{q}$ (red line) and the spot size $\sigma_\mathrm{r}$ of the radiation at focus (blue line). (d) On-axis $E_x$ (orange line) and central frequency (blue line) at $t=106$ fs. The blue dashed line denotes the theoretical chirp curve of the radiation, and the red line denotes the result with the QED module enabled. The inset shows the dependence of radiation peak intensity $I_0$ at $t=106$ fs on the initial beam density $n_\mathrm{b}$. The squares are the simulation results and the line is a quadratic fitting. \label{fig:3D}}
\end{figure}

Figures \ref{fig:3D}(a) and (b) show the electron beam density and the radiation field distributions at $t=53$ fs and $t=106$ fs, respectively. The blue line in each panel denotes the on-axis profile of the electron beam density, which is similar to that in the 1D case. The beam expands transversely due to the space-charge repulsion between electrons, which reduces the quasiparticle density $n_\mathrm{q}$ as shown by the red line in Fig. \ref{fig:3D}(c). The transverse Gaussian distribution of the laser pulse results in a curved density surface \cite{VincentiPhysRevLett.123.105001} that can focus the radiation. As the laser-beam interaction proceeds, the beam front surface becomes increasingly curved, as described by $\xi^\mathrm{q}(r)\approx -\frac{3[(1+a_\mathrm{L}^2(r))ct]^{1/3}}{2(-k)^{2/3}}$. The reduction in the radius of curvature of the beam surface, together with the decreasing radiation wavelength, results in a smaller focal spot size of the radiation, as evidenced by comparing the radiation profiles in Figs. \ref{fig:3D}(a) and (b). The blue line in Fig. \ref{fig:3D}(c) depicts the evolution of the focal spot size, which achieves 50 nm spot size at $t=106$ fs.

The radiation pulse has a peak power of $1.2$ GW and a pulse energy of $0.9~\micro$J. It is chirped, as shown by the on-axis $E_x$ [orange line in Fig. \ref{fig:3D}(d)] and the shift of its central wavelength from 105 to 14 nm (blue solid line). The theoretical chirp is given by $\omega(\xi)=\omega_\mathrm{L}\frac{16k^2\xi^2}{9(1+a_\mathrm{L}^2)}$ (blue dashed line), which agrees well with the simulation. If the pulse is fully compressed, this large bandwidth can support an ultrashort pulse duration of $\sim10$ as and a peak power of $\sim100$ GW. The inset in Fig. \ref{fig:3D}(d) presents the dependence of the radiation peak intensity $I_0$ at $t=106$ fs on the initial beam density $n_\mathrm{b}$, which illustrates a typical quadratic scaling of superradiance.

\section*{Discussion}
Quasiparticle formation is not restricted to circular polarization, but remains robust under linearly polarized driving through the same sheet-crossing mechanism. Although the longitudinal motion exhibits rapid oscillations, the cycle-averaged dynamics and the associated chirped radiation resemble those in the CP case, underscoring the generality of the underlying mechanism. Beam emittance introduces an initial transverse momentum ($p_{\perp0}$) that modifies the longitudinal velocity as $v_z\approx \left[ 1-\frac{1+a_\mathrm{L}^2+(p_{\perp0}/m_\mathrm{e}c)^2}{2\gamma_0^2} \right]c$. For $p_{\perp0}\ll \gamma_0m_\mathrm{e}c$, this is equivalent to an electron with zero transverse momentum but a modified initial energy of $\gamma_0\sqrt{\frac{1+a_\mathrm{L}^2}{1+a_\mathrm{L}^2+ (p_{\perp0}/m_\mathrm{e}c)^2}}$. This transverse momentum spread effectively induces an additional energy spread of $\frac{\gamma_0}{2(1+a_\mathrm{L}^2)} \Delta\left( \frac{p_{\perp0}}{m_\mathrm{e}c} \right)^2$. With our operating parameters ($\gamma_0 \sim 60$ and an angular spread $\Delta \left( \frac{p_{\perp0}}{\gamma_0 m_\mathrm{e}c} \right)$ of $\sim 1$ mrad), this effective spread is $\sim 10^{-3}$, which is negligible against the energy spread ($\sigma_{\gamma_0} \sim 0.1$). Additionally, transverse beam expansion is dominated by space-charge effects, exceeding emittance-driven expansion by orders of magnitude, especially for lower-energy slices (see Supplemental Material). Ultimately, the impact of beam emittance on quasiparticle dynamics is negligible in our configuration.

\begin{figure}[!htbp]
\centering
\includegraphics[width=0.5\linewidth]{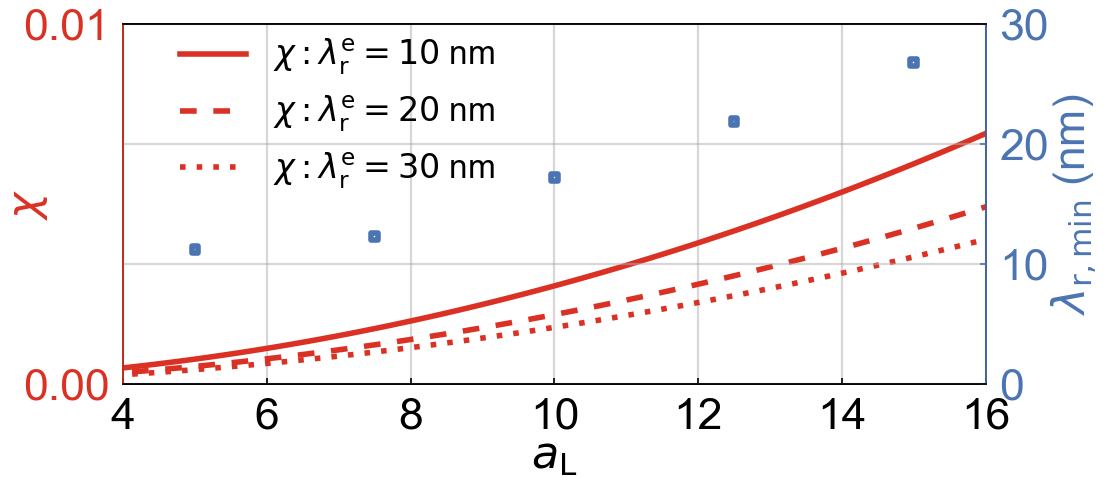}
\caption{ The dependence of the quantum parameter $\chi$ at $\lambda_\mathrm{r}^\mathrm{e}=10,20,30$ nm (red lines), and the shortest coherent radiation wavelength $\lambda_\mathrm{r,min}$ (blue dots) on the laser amplitude $a_\mathrm{L}$. The electron beam parameters are identical to those in Fig. \ref{fig:concept}(c). }\label{fig:aL-scan}
\end{figure}

Furthermore, we evaluate the influence of quantum electrodynamics (QED) effects on the quasiparticle dynamics. While the mean electron energy loss induced by classical radiation reaction only slightly shifts the radiation frequency \cite{classical_RR_calculation,RR-theory}, stochastic photon emission can increase the slice energy spread \cite{QED_energy_spread,QED-review} and broaden the quasiparticle profile, thereby potentially suppressing coherent emission at short wavelengths. Importantly, this limitation can be effectively circumvented by optimizing the parameter space. Since the single particle radiation wavelength is given by $\lambda_\mathrm{r}^\mathrm{e}=\frac{\lambda_\mathrm{L}}{4\gamma^2}(1+a_\mathrm{L}^2)$, we can achieve the same $\lambda_\mathrm{r}^\mathrm{e}$ by simultaneously reducing $a_\mathrm{L}$ and the electron energy. This adjustment significantly lowers the quantum parameter $\chi \propto a_\mathrm{L}\gamma$, thereby mitigating stochastic broadening. To systematically quantify this trend, we conduct a parametric scan over $a_\mathrm{L}$ using 1D PIC simulations incorporating QED effects. Figure \ref{fig:aL-scan} shows the shortest radiation wavelength $\lambda_\mathrm{r,min}$ extracted from the Wigner distribution (blue dots) alongside the calculated $\chi$ values for various $\lambda_\mathrm{r}^\mathrm{e}$ (red lines). As $a_\mathrm{L}$ decreases, $\chi$ drops substantially, alleviating the QED-induced broadening of the quasiparticle and gradually restoring the short-wavelength coherent radiation. Notably, when $a_\mathrm{L} \lesssim 7.5$, coherent emission down to $\sim 10$~nm is successfully generated, revealing that quantum effects become negligible under the considered parameters. Finally, Fig. \ref{fig:3D}(d) presents the on-axis central wavelength distribution for the $a_\mathrm{L}=10$ case with QED effects included (red line). Here, QED effects predominantly impact the short-wavelength spectrum, shifting the cutoff wavelength from 14~nm to 19~nm.

The first-class quasiparticle can be investigated experimentally at all-optical Thomson scattering facilities based on laser-driven PBAs \cite{PhysRevLett.96.014802, ta2012all, powers2014quasi, PhysRevLett.114.195003, yan2017high, PhysRevX.8.011020, PhysRevX.8.031004, Ma2020MRE, mirzaie2024all, matheron2024comptonphotonsgevscale, hpl2025sjtu-ts} since PBAs can provide the needed chirped electron beams. Meanwhile, the monoenergetic electron beams used in the formation of the second-class quasiparticle can be accessible at Thomson scattering light sources based on conventional accelerators \cite{Schoenlein1996Science, Gibson2004PLEIADES, WELLER2009HIgammaS, TTX2013RSI, FukudaIPAC2016-TUPOW046, PhysRevAccelBeams.22.053403, Wang2022SLEGS, SAMSAM2024STAR,  PhysRevAccelBeams.28.023401, Gunther2023, Lyncean2017, LumitronTechnologies}. 

In conclusion, we have shown that generalized superradiance can be used to convert incoherent Thomson scattering facilities into coherent radiation sources. Two distinct categories of quasiparticles emerge: one formed via electron sheet crossing and the other via electron aggregation, and they respectively generate a chirped pulse with tens of nm wavelength and monochromatic THz radiation. This feasible scheme can be implemented in Thomson scattering facilities, extending their capability for coherent radiation generation and opening up broad applications across multiple disciplines.






 

\section*{Methods}
\textbf{Simulation setup}\\
\noindent The simulations are carried out using the fully relativistic, multi-dimensional electromagnetic particle-in-cell (PIC) code OSIRIS \cite{osiris}. Simulation boxes moving with the speed of light $c$ in vacuum are used. The basic settings for the simulations are listed in Table. \ref{tab:1}.

\begin{table}[ht]
\centering
\caption{\label{tab:1}%
\color{black}{Basic settings of PIC simulations.}}
\begin{tabular}{ccccc}
\hline
\hline
\textrm{Figure}&
\textrm{Box size ($\micro\meter$)}&
\textrm{Grid numbers}&
\textrm{Timestep (as)}&
\textrm{Particle per cell}\\
\hline
 Fig. 2(a) & 5.3 & 4000 & 4.4 & 1000 \\
 Fig. 2(b) & 111.3 & 21000 & 17.7 & 200 \\
 Fig. 4 & 3.2 $\times$ 19.1 $\times$ 19.1 & 2400 $\times$ 1200 $\times$ 1200 & 2.2 & 1 \\
\hline
\hline
\end{tabular}
\end{table}

In the one-dimensional (1D) simulations for the first class quasiparticles, the electric field of the laser pulse starts with a profile as $10\left( \frac{\tau}{35~\femto\second}\right)^3-15\left( \frac{\tau}{35~\femto\second}\right)^4+6\left( \frac{\tau}{35~\femto\second}\right)^5$ where $\tau \in (0,35)~\femto\second$, followed by a constant envelope with $a_\mathrm{L}=15$. The electron beam is $2.1~\micro\meter$ long and has a flattop longitudinal profile. The beam has a negative energy chirp $k\equiv\frac{d\gamma_0}{d\xi_0}=-37.6~\micro\meter^{-1}$ with $\gamma_0\in [20,100]$. The slice energy spread of the beam is $\sigma_{\gamma_0}=0.1$. In the 1D simulations for the second class quasiparticles, the envelope of the normalized vector potential of the flying focus laser is set as $a_\mathrm{L}\mathrm{exp}\left[ -\frac{2\mathrm{ln}2(z - v_\mathrm{f}t)^2}{c^2\tau_\mathrm{FWHM}^2} \right]$, where $\tau_\mathrm{FWHM}=0.42$ fs and $v_\mathrm{f}=-0.95c$ (see Sec. II for more details). The electron beam is $106~\micro\meter$ long and has a flattop longitudinal profile. The beam is monoenergetic with $\gamma_0=10$. The slice energy spread of the beam is $\sigma_{\gamma_0}=0.1$ and the initial transverse momentum spread follows a Gaussian distribution with an rms width of $0.1m_ec$.

In the 3D simulations, a recently developed advanced Maxwell solver \cite{xu-solver} is used to suppress the numerical fields \cite{xu2013numerical} and model the interactions between dense electrons and intense electromagnetic fields with high-fidelity. The electron beam has a root-mean-square (rms) spot size of $1.0~\micro\meter$ and a flattop longitudinal envelope with a beam length of $1.6~\micro\meter$. The beam has a negative energy chirp $k\equiv\frac{d\gamma_0}{d\xi_0}=-56.4~\micro\meter^{-1}$ with $\gamma_0\in [20,110]$. Other parameters are the same as the 1D cases. The foci of the flying focus laser travels at $v_\mathrm{f}=0.95c$ with a rms spot size of $4.2~\micro\meter$. All lasers employed in the simulations are circularly polarized with $a_\mathrm{L} = 15$.

\hspace*{\fill}

\noindent\textbf{Numerical modeling of the flying focus laser field}\\
\noindent The 1D flying focus laser is modeled using a vector potential $\bm{A}$ that captures its spatiotemporal shape as $A_x=-A(z-v_\mathrm{f}t)\cos[k_\mathrm{L}(z+ct)]$ and $A_y=A(z-v_\mathrm{f}t)\sin[k_\mathrm{L}(z+ct)]$ \cite{2023jacob-flyfoc}. By assuming a Gaussian temporal envelope $A(z-v_\mathrm{f}t)=a_\mathrm{L}\mathrm{exp}\left[-\frac{2\mathrm{ln}2(z-v_\mathrm{f}t)^2}{c^2\tau_\mathrm{FWHM}^2}\right]$, the electromagnetic fields of a 1D flying focus laser is given by
\begin{align}
    E_x&=A(z-v_\mathrm{f}t)[4\mathrm{ln}2 v_\mathrm{f}\frac{z-v_\mathrm{f}t}{c^2\tau_\mathrm{FWHM}^2}\cos(k_\mathrm{L}(z+ct))-ck_\mathrm{L}\sin(k_\mathrm{L}(z+ct))] ,  \\
    E_y&=A(z-v_\mathrm{f}t)[-4\mathrm{ln}2v_\mathrm{f}\frac{z-v_\mathrm{f}t}{c^2\tau_\mathrm{FWHM}^2}\sin(k_\mathrm{L}(z+ct))-ck_\mathrm{L}\cos(k_\mathrm{L}(z+ct))] ,  \\
    B_x&=A(z-v_\mathrm{f}t)[4\mathrm{ln}2\frac{z-v_\mathrm{f}t}{c^2\tau_\mathrm{FWHM}^2}\sin(k_\mathrm{L}(z+ct))-k_\mathrm{L}\cos(k_\mathrm{L}(z+ct))] ,  \\
    B_y&=A(z-v_\mathrm{f}t)[4\mathrm{ln}2\frac{z-v_\mathrm{f}t}{c^2\tau_\mathrm{FWHM}^2}\cos(k_\mathrm{L}(z+ct))+k_\mathrm{L}\sin(k_\mathrm{L}(z+ct))] .
\end{align}

In the 3D geometry, each longitudinal slice $j$ of the flying focus pulse is focused at a different location along the propagation axis, and the transverse electric field of the flying focus laser is given by \cite{2023jacob-flyfoc}
\begin{equation}
    \bm{E}_\perp(\bm{x}_\perp,z,t)=\sum_{j=1}^N \frac{1}{2}A_0 B_j(\zeta)C_j(\bm{x}_\perp,z)e^{ik_\mathrm{L}\zeta}\bm{\epsilon}+c.c. ,
\end{equation}
where $\zeta=z+ct$ denotes the coordinate of a slice $j$, $\bm{\epsilon}$ is the polarization vector, $A_0$ is chosen to produce a desired amplitude $a_\mathrm{L}=15$, $B_j(\zeta)$ is the temporal envelope and $C_j(\bm{x}_\perp,z)$ is the transverse profile. 

In our simulations, $C_j(\bm{x}_\perp,z)$ is a Gaussian transverse profile. Each slice $\zeta$ is assigned a focal point $s_0(\zeta)$. The velocity of the intensity peak $v_\mathrm{f}$ is controlled by setting $s_0=\alpha\zeta$, where $\alpha=\frac{v_\mathrm{f}}{v_\mathrm{f}+c}$ for a backward pulse ($\zeta=z+ct$) and $\alpha=\frac{v_\mathrm{f}}{v_\mathrm{f}-c}$ for a forward pulse ($\zeta=z-ct$). Moreover, for a flying focus laser with a constant temporal envelope, i.e. $a_\mathrm{L}=15$, its on-axis intensity exhibits a peak traveling at $v_\mathrm{f}$, with a FWHM duration of $\tau_\mathrm{FWHM}=2\sqrt{3}(1+v_\mathrm{f}/c)z_\mathrm{R}$ in 2D and $\tau_\mathrm{FWHM}=2(1+v_\mathrm{f}/c)z_\mathrm{R}$ in 3D, where $z_\mathrm{R}=\frac{\pi w_0^2}{\lambda_\mathrm{L}}$ is the Rayleigh length.

\section*{Data availability}
The data that support the findings of this study are available from the corresponding author upon reasonable request.

\section*{Code availability}
All computer codes supporting the findings of this study are fully documented within the paper and its references. Reasonable additional inquiries about the codes should be directed to the corresponding author.

\bibliography{ref}

\section*{Acknowledgements}

This work was supported by the Fundamental and Interdisciplinary Disciplines Breakthrough Plan of the Ministry of Education of China-JYB2025XDXM204, the National Grand Instrument Project (No. 2019YFF01014400), the National Natural Science Foundation of China (NSFC) (No. 12375147), Guangdong Provincial Science and Technology Plan Project (2021B0909050006), the Fundamental Research Funds for the Central Universities, Peking University. The simulations were supported by the High-performance Computing Platform of Peking University. We would like to thank Jacob R. Pierce for useful discussions.

\section*{Author contributions}
Q. M., Y. X., Z. W., L. L., Z. Y., X. X. and X. Y. contributed extensively to the work presented in this paper.

\section*{Competing interests}
The authors declare no competing interests.

\section*{Additional information}



\end{document}